\documentclass[twocolumn,english,prl]{revtex4}
\usepackage[T1]{fontenc}
\usepackage[latin1]{inputenc}
\usepackage{graphicx}

\makeatletter

\providecommand{\LyX}{L\kern-.1667em\lower.25em\hbox{Y}\kern-.125emX\@}

\usepackage{babel}
\makeatother
\begin{document}

\title{Non-exponential Dissipation in a Lossy Elastodynamic Billiard, Comparison
with Porter-Thomas and Random Matrix Predictions}

\author{Oleg I. Lobkis, Igor S. Rozhkov and Richard L. Weaver}

\affiliation{Theoretical \& Applied Mechanics 104 So Wright Street University
of Illinois Urbana Illinois 61801}

\begin{abstract}
We study the dissipation of diffuse ultrasonic energy in a reverberant
body coupled to a waveguide, an analog for a mesoscopic electron in
a quantum dot. A simple model predicts a Porter-Thomas like distribution
of level widths and corresponding nonexponential dissipation, a behavior
largely confirmed by measurements. For the case of fully open channels,
however, measurements deviate from this model to a statistically significant
degree. A random matrix supersymmetric calculation is found to accurately
model the observed behaviors at all coupling strengths.
\end{abstract}
\maketitle
A random matrix model of wave scattering from an ideal lead, off a
chaotic region, has been employed in nuclear theory for decades {[}1-6{]}.
It originated in the theory of nuclear reactions and is widely used
in the analysis of compound nuclei, mesoscopic quantum dots and microwave
cavities {[}1-6{]}. It has provided a model of intrinsic loss mechanisms
in classical wave chaotic systems, with implications for microwave
cavities and reverberant ultrasonics. In particular it has been shown
that dissipation is not necessarily exponential in time, and made
specific predictions for how that decay should behave.

A simple argument that relates such behavior to the distribution of
resonance widths leads to a non-exponential decay law under the assumption
of a Porter-Thomas (chi-square) width distribution \cite{key-5,key-15}.
That distribution follows from first order perturbation theory on
a system with Gaussian mode shape statistics and a finite number of
loss channels. A more general argument may be constructed by the information-theoretic
approach \cite{key-2} or the random Hamiltonian approach \cite{key-4,key-8}.
These also predict non-exponential decay and corresponding delay time
distributions, but with different details \cite{key-4,key-5,key-11}.
Acoustic \cite{key-15} and microwave \cite{key-14} measurements
have confirmed non-exponential decays. Critical comparisons with the
models have not been undertaken.

In this Letter we address diffuse energy decay in an open acoustical
system. While non-exponential decays have long been observed, and
modeled by means of Porter-Thomas distributions of resonance widths,
deviations from those predictions have not yet been detected. Indeed,
it has not yet been clear whether measurements can distinguish between
the predictions of the simple argument and those of random matrix
theory (RMT) {[}3--9{]}. It is therefore interesting to compare both
the RMT predictions \cite{key-8} and the simple argument predictions
\cite{key-15} with measurements conducted on an experimental realization
of a chaotic billiard attached to a waveguide. In most cases we find
that a Porter-Thomas model does an adequate job of fitting observed
decay profiles. In the case of strongly coupled channels, however,
a full RMT prediction is superior. The difference is small, but statistically
significant.%
\begin{figure}[htbp]
\begin{center}\includegraphics{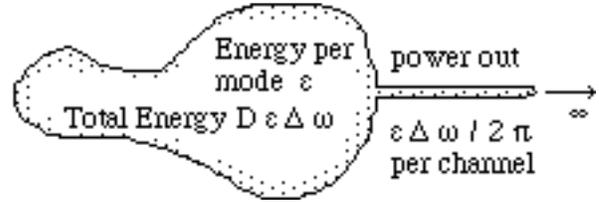}\end{center}

\caption{A diffuse wavefield radiates into a waveguide.}
\end{figure}

Consider a reverberant body as in Fig. 1 coupled to a waveguide. The
body has mean diffuse energy density (per mode) of $\varepsilon $.
The total energy in a band of width $\Delta \omega $ is $\varepsilon D\Delta \omega $,
where the body's modal density $D=\partial N_{Body}/\partial \omega $.
An attached waveguide has mean energy per outgoing mode no greater
than $\varepsilon $. Incoming modes have no energy. Each guided mode
(i.e. each channel) therefore carries power at a rate no greater than
$(d\Delta \omega )(\varepsilon /2)v_{g}$, where $d$ is the lineal
modal density per length in the channel, $d=\partial N_{channel}/\partial \omega \partial L=1/\pi v_{g}$
and $v_{g}$ is the group velocity of that mode. The factor ($1/2$)
is due to neglect of the incoming modes. Thus each channel conducts
outgoing power $\Pi \leq (1/2\pi )\varepsilon \Delta \omega $, independent
of dispersion in the channel. At maximal coupling each channel contributes
the same mean partial width, an energy decay rate of $\Pi /\varepsilon D\Delta \omega =1/[2\pi \partial N_{Body}/\partial \omega ]\equiv 1/t_{Heisenberg}$.
This picture does not apply to individual normal modes of the body,
but only to the mean. Those normal modes of the body which overlap
well with the waveguide will dissipate rapidly; those which overlap
poorly will do so slowly. While the average level width (and early
time decay rate) may well correspond with the above estimate, the
less strongly dissipated modes will eventually dominate a transient
decaying field; the apparent dissipation rate will appear to diminish.

Modeling the diffuse field as a superposition of independent real
normal modes with Gaussian statistics (this is not correct, lossy
systems generically have complex eigenmodes), each with a proper decay
rate given by first order perturbation theory, leads to a chisquare-like
distribution of modal decay rates, and a net transient energy decay\begin{equation}
E(t)=E_{0}\prod _{i=1}^{M}\left(1+2\sigma _{i}t\right)^{-1/2}\label{et1}\end{equation}
where $M$ is the number of open outgoing channels, and $\sigma _{i}$
is the decay rate through the $i$th channel, $\sigma _{i}\leq 1/t_{H}$.
In certain limits, it reduces $E(t)\approx E_{0}\exp \left(-\sum _{i}\sigma _{i}t\right)$.

In the special case of equipotent channels, $\sigma _{i}=\sigma /M$,
one recovers the simplest Porter-Thomas model \cite{key-15},\begin{equation}
E(t)=E_{0}\left(1+2\sigma t/M\right)^{-M/2}\label{et2}\end{equation}

This form for the dissipation of ultrasonic energy density has been
confirmed phenomenologically by taking $E_{0}$, $M$, and $\sigma $
to be adjustable parameters. Observed profiles $E(t)$ have been found
to fit remarkably well \cite{key-15}. A particularly noteworthy case
is that of \cite{key-14} in which the case of three weakly coupled
($\sigma _{i}\ll 1/t_{H}$) channels was successfully fit to Eq. (\ref{et1}).
In many of these fits it has not been clear whether the recovered
parameter $M$ is meaningful, i.e. whether it does indeed correspond
to a discrete number of equipotent effective loss channels.

A better theory, beyond first order perturbation, is provided by considering
a random matrix model \cite{key-3,key-4} for which level width distributions
are known to be non-chisquare \cite{key-4}. Here the dynamics is
governed by a (GOE) Gaussian random Hamiltonian, consistent with the
assumed chaotic ray trajectories of the body, plus an anti-Hermitian
part corresponding to a discrete number $M$ of outgoing channels.
Supersymmetric techniques \cite{key-4,key-8} allow construction of
various averages. Amongst the more easily constructed is $E(t)=\left|G_{ij}\left(t\right)\right|^{2}$,
the mean square response at a site $j$ distinct from the site $i$
of the source, each distinct from any dissipative sites. $G_{ij}\left(t\right)$
is the time-domain Green's function. On inverse Fourier transforming
the result of a supersymmetric calculation \cite{key-8} one finds:\begin{eqnarray}
 & E(\tau )\sim \int _{1}^{\infty }\int _{1}^{\infty }d\lambda _{1}d\lambda _{2}\Pi \left(\tau ,\lambda _{1},\lambda _{2}\right)f\left(\tau ,\lambda _{1},\lambda _{2}\right) & \nonumber \\
 & \times \frac{\theta \left(\lambda _{1}\lambda _{2}-2\tau +1\right)\theta \left(2\tau -\lambda _{1}\lambda _{2}+1\right)\left(1-\left(2\tau +\lambda _{1}\lambda _{2}\right)^{2}\right)}{\left(\lambda _{1}^{2}+\lambda _{2}^{2}+\left(2\tau +\lambda _{1}\lambda _{2}\right)^{2}-2\lambda _{1}\lambda _{2}\left(2\tau +\lambda _{1}\lambda _{2}\right)-1\right)^{2}} & \label{eoftau}
\end{eqnarray}
where, $\tau =t/t_{H}$, $M$ is the number of channels, each characterized
by a coupling parameter $g_{i}\geq 1$, $f=\left(\lambda _{1}^{2}-1\right)\lambda _{2}^{2}+\left(\lambda _{2}^{2}-1\right)\lambda _{1}^{2}+1-\left(2\tau +\lambda _{1}\lambda _{2}\right)^{2}$,
$\Pi =\prod _{i=1}^{M}\left(g_{i}+2\tau +\lambda _{1}\lambda _{2}\right)\left(g_{i}^{2}+2g_{i}\lambda _{1}\lambda _{2}+\lambda _{1}^{2}+\lambda _{2}^{2}-1\right)^{-1/2}$.
For a large number of weak ($g_{i}\gg 1$) channels, this reduces
to: $E(t)=E_{0}\exp \left(-\tau \sum 2/\left(g_{i}+1\right)\right)$,
in agreement with the naive model. At finite $M$ the two models differ
slightly. A comparison at $M=4$, $g_{i}=1.5$, $\sigma _{i}=2/\left(g_{i}+1\right)t_{H}=4/5t_{H}$
for all $i$, is shown in Fig. 2. The naive model overestimates curvature.%
\begin{figure}[htbp]
\begin{center}\includegraphics[  width=3.375in,
  keepaspectratio]{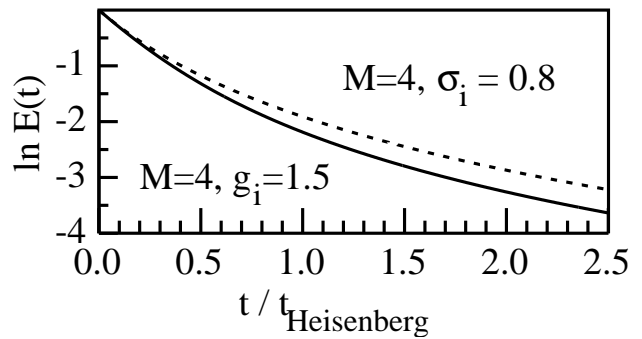}\end{center}

\caption{A comparison of the predictions of the Porter-Thomas model (dashed
line, Eq. (\ref{et1})) for transient decay with that of a full supersymmetric
calculation (solid line, Eq. (\ref{eoftau})).}
\end{figure}

We study the aluminum body (volume $561\, cm^{3}$, free surface $451\, cm^{2}$)
pictured in Fig. 3. Non parallel faces and defocusing surfaces enhance
ray-chaos. After preliminary baseline measurements, it was welded
to an aluminum wire ($1100$ alloy, $3.18\, mm$ diameter, $3\, m$
length). Tests were carried out with the spiral part submerged in
a water bath. Attenuation in the water assures negligible reflected
energy and thus the presence of only outgoing waves. This was confirmed
by separate measurements. The guided elastic waves of a circular waveguide
are described by the Pochammer dispersion relation \cite{key-16}.
All modes with azimuthal number $n>0$ are two-fold degenerate. At
low frequency, below the first cutoff at $580\, kHz$, there are $M=4$
propagating guided waves. Two are flexural ($n=1$), one is extensional
($n=0$) and one torsional ($n=0$). A calculation of cutoff frequencies
gives the number of open channels. The strength with which these channels
are coupled is not known apriori. Measurement of reflections of waves
incident from the wire onto the block indicate that the coupling is
good; reflections are generally weak. Dispersion and the possibility
of mode conversion complicate any attempt to be more quantitative.%
\begin{figure}[htbp]
\begin{center}\includegraphics{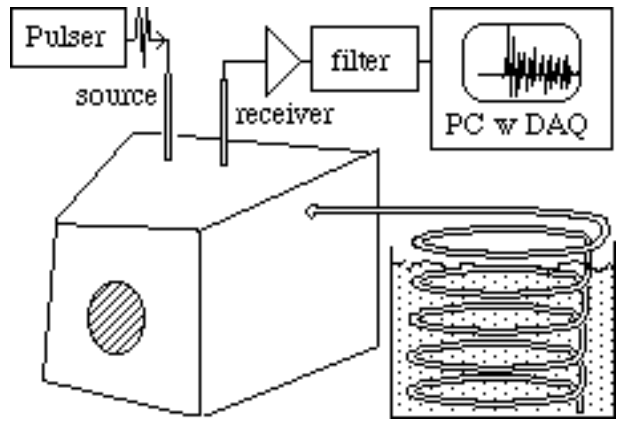}\end{center}

\caption{Sketch of measurement system.}
\end{figure}

Profiles $E(t)$ were constructed both before and after the waveguide
was attached. For each case, piezoelectric pulses of negligible duration
were applied to pin-transducers in light oil contact with the body,
as in Fig. 3. Waveforms of durations of up to $100\, msec$, and bandwidths
to $2\, MHz$ were recorded at a digitization rate of $5\, MSa/sec$.
A low pass filter with a cutoff of $2.25\, MHz$ prevented aliasing.
Repetition averaging improved signal to noise ratios and extended
the system's dynamic range. The resulting waveforms were time-windowed
into $62$ successive $1.64\, msec$ sections with tapered edges.
Each windowed waveform was Fourier transformed and squared and integrated
over rectangular bins of width $25$ or $50\, kHz$. The result was
an array of spectral energy densities versus time for each of several
narrow frequency bands. On repeating this for $16$ distinct source
and receiver positions an average $E(t)$ was constructed for each
band. A typical profile is seen in Fig. 4. As expected, the waveguide
has augmented the decay rate.%
\begin{figure}[htbp]
\begin{center}\includegraphics{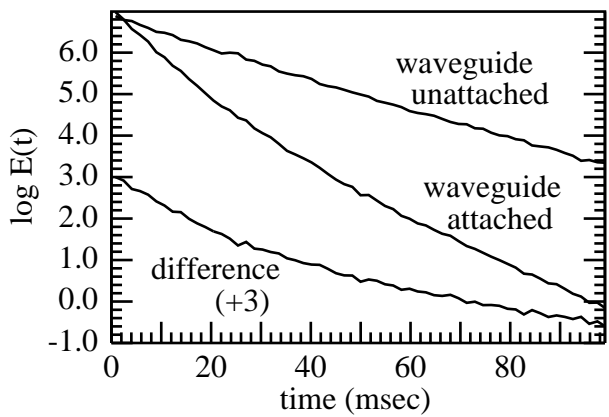}\end{center}

\caption{$\ln E(t)$ averaged over $16$ distinct positions of source and
receiver in a band between $225$ to $275\, kHz$. The Heisenberg
time is $34\, msec$.}
\end{figure}

The upper curve is the reference case without the waveguide; it shows
a decay which is very nearly exponential, i.e. consistent with intrinsic
decay mechanisms being widely distributed and corresponding to a large
number of weakly coupled dissipative channels. As in \cite{key-15},
the behavior fits well to Eq. (\ref{et2}); chisquares are excellent.
While the parameter M extracted from that fit is perhaps not meaningful,
we do take Eq.(\ref{et2}) as a valid way to smooth the reference
data. The smoothed reference $\ln E(t)$ is then subtracted from the
measured $\ln E(t)$ in the waveguide-attached case to give a difference
$\ln E$, what we would have measured in the attached case if our
reference block had been non-dissipative. This profile shows substantial
curvature, consistent with a hypothesis of a small number of outgoing
channels.

The observed difference $\ln E$'s were then themselves fit to Eq.
(\ref{et2}). Figs. 5 and 6 show the parameters $\sigma $ and $M$
extracted from those fits, and compares them with expectations based
on assuming perfect coupling: $M_{Pochammer}=$ number of propagating
modes; $\sigma =M_{Pochammer}/t_{H}$. That the $\sigma $ of the
difference $\ln E$ is not in excess of this theory, even at high
frequencies, is an indication that the welding process has not significantly
increased intrinsic absorption. The correspondence is remarkably good
for such a simple theory; $M$ and $\sigma $ show the predicted features,
in particular those associated with the onsets of new guided modes
at $580$, $800$ and $1000\, kHz$. At low frequencies, the correspondence
is poor; the model assumption of perfect coupling is incorrect.%
\begin{figure}[htbp]
\begin{center}\includegraphics{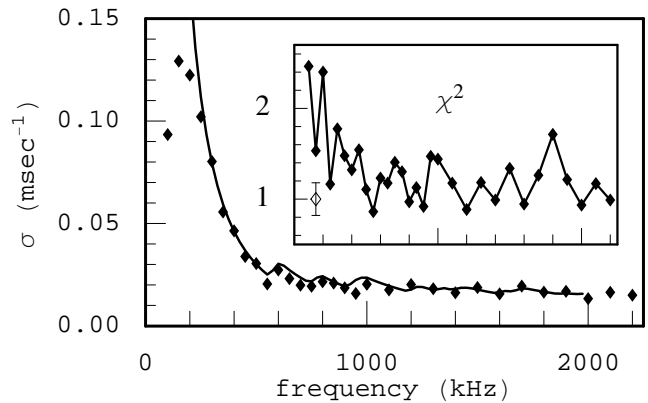}\end{center}

\caption{Decay rates $\sigma $ as recovered from a fit of the difference
$\ln E(t)$'s to Eq. (\ref{et2}) in each of several narrow frequency
bands (data points) are compared to the prediction: $\sigma =M_{Pochammer}/t_{H}$
(solid line).}
\end{figure}

\begin{figure}[htbp]
\begin{center}\includegraphics[  width=3.375in,
  keepaspectratio]{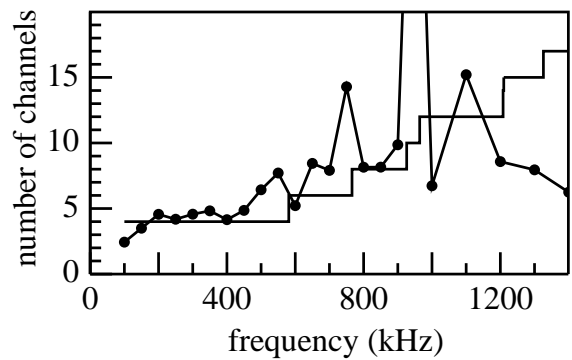}\end{center}

\caption{$M$ values as recovered from fits to the difference $\ln E(t)$'s
in each of several narrow frequency bands are compared to expectations
based on the cutoff frequencies of the Pochammer dispersion relation.}
\end{figure}

The reduced chisquares of these fits are shown in the inset. At high
frequencies they are within one standard deviation of the expected
value of unity. Plots of the residuals show that fluctuations are
not systematic. We conclude that the higher-frequency data is consistent
with Eq. (\ref{et2}); small curvatures are well modeled by a single
phenomenological parameter $M_{effective}$; the high frequency data
does not permit conclusions in regard to the relative virtues of the
various theories. At low frequency the chisquares exceed unity. This
is in part due to the inadequacy of our spatial averaging there; different
transducer positions are often within a wavelength and are correlated.
It is in part also due to some systematic deviations that indicate
a need for a better theory; unphysical values of $M$ (Fig. 6) are
a further indication of that need.%
\begin{figure}[htbp]
\begin{center}\includegraphics{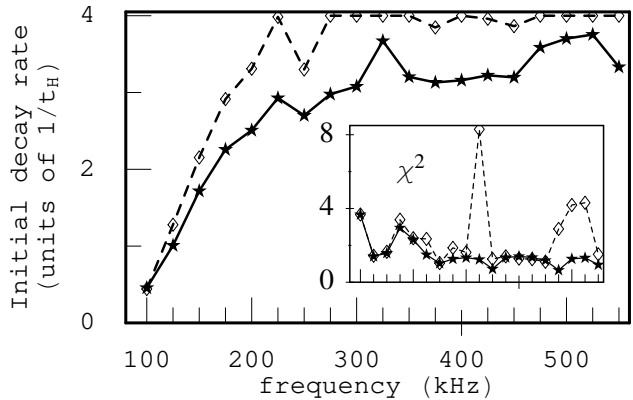}\end{center}

\caption{Values for $t_{H}\sum \sigma _{i}$ (dashed line) and for $\sum 2/\left(1+g_{i}\right)$
(solid line) from fits of Eq. (\ref{et1}) and Eq. (\ref{eoftau})
to the difference $\ln E(t)$ profiles. The chisquares of the fits
indicate Eq. (\ref{et1})'s inability to fit the data in those places
where coupling is strong.}
\end{figure}
\begin{figure}[htbp]
\begin{center}\includegraphics[  width=3.375in,
  keepaspectratio]{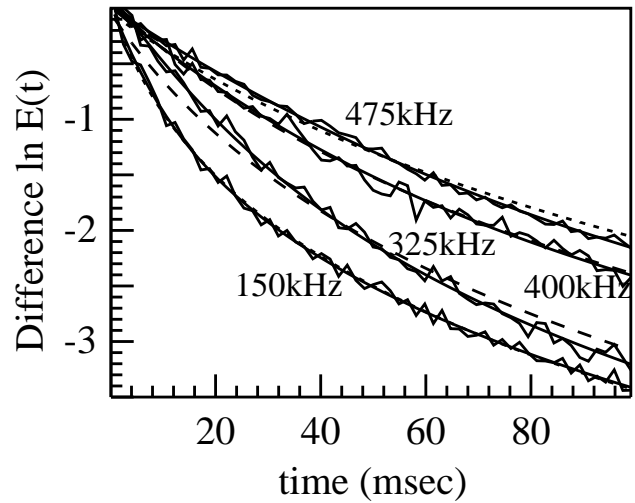}\end{center}

\caption{Fits of the supersymmetric profile (Eq. (\ref{eoftau})) (smooth
solid lines) and the Porter-Thomas model (Eq. (\ref{et1})) (dashed
lines) to the difference $\ln E(t)$'s (irregular solid lines) in
representative frequency bins. At $150\, kHz$ each model fits well:
Eq. (\ref{et1}) calls for $\sigma _{i}=\left\{ 1,\, 1,\, 0.07686,\, 0.07686\right\} /t_{H}$
(the flexural modes are especially weakly coupled at such long wavelength,
hence their small $\sigma $'s.) Eq. (\ref{eoftau}) calls for $2/\left(1+g_{i}\right)=\left\{ 0.5954,\, 1,\, 0.0614,\, 0.0614\right\} $.
At $475\, kHz$, the fits call for $\sigma _{i}=1/t_{H}$ and $2/\left(1+g_{i}\right)=0.9$
for all $i$, }
\end{figure}

At low frequencies ($<580\, kHz$), where there are only four open
channels, we attempt a fit to the richer theories ($1$) and ($3$)
by fixing $M$ at four and adjusting the coupling strengths. We take
the two flexural waves to have identical coupling (the weld is axisymmetric);
set $g_{3}=g_{4}$, $\sigma _{3}=\sigma _{4}$, and adjust only $E_{0}$
and three values of $g_{i}\geq 1$ (or $\sigma _{i}\leq 1/t_{H}$).
The fits' values for total loss coefficient ($t_{H}\sum \sigma _{i}\leq 4$
and $\sum 2/\left(1+g_{i}\right)\leq 4$) are shown in Fig. 7. Representative
plots of the data and fits are shown in Fig. 8. In cases with weak
coupling, both models do well. At higher frequencies where coupling
is efficient (all $g$'s close to unity), and first order perturbation
theory is invalid, there are significant discrepancies. At several
of the frequencies only the full RMT calculation Eq. (\ref{eoftau})
is statistically acceptable. At a few other frequencies where Eq.
(\ref{et1}) gives an acceptably low chisquare, its call for all four
channels to be perfectly coupled, $\sigma _{i}=1/t_{H}$, is implausible.
The fits at these several frequencies may be the first demonstration
of non-Porter-Thomas like level width distributions; it is also evidence
for the superior applicability of a full RMT calculation over that
of simpler theories. This has implications for RMT and wave chaos,
for mesoscopics, nuclear physics, microwave physics, diffuse field
ultrasonics, and also for structural acoustics where modeling of losses,
both intrinsic and radiative \cite{key-17,key-18}, is gathering increased
attention.

This work was supported by NSF grant CMS-0201346.


\begin{thebibliography}{10}
\bibitem{key-1}C. Mahaux, H.A. Weidenmüller, \emph{Shell-Model Approach to Nuclear
Reactions}. (North-Holland, Amsterdam, 1969).
\bibitem{key-2}P.A. Mello, H.U. Baranger, Waves in Random Media 9, 105 (1999); C.
W. J. Beenakker, Rev. Mod. Phys. 69, 731 (1997)
\bibitem{key-3}T. Guhr, A. Müller-Groeling, H.A. Weidenmüller, Phys. Rep. 299, 189
(1998); H.-J. Stöckmann. \emph{Quantum chaos. An Introduction}. (Cambridge
University Press, 1999); Y. Alhassid, Rev. Mod. Phys. 72, 895 (2000).
\bibitem{key-4}Yan V. Fyodorov, H.-J. Sommers, J. Math. Phys. 38, 1918 (1997).
\bibitem{key-8}J. M. Verbaarschot, H.A. Weidenmüller and M.R. Zirnbauer, Phys. Rep.
129, 367 (1985); J. M. Verbaarschot, Ann. of Physics, 168, 138 (1986);
K. Efetov. \emph{Supersymmetry in disorder and chaos} (Cambridge University
Press, 1997).
\bibitem{key-5}F.-M. Dittes, Phys. Rep. 339, 215 (2000).
\bibitem{key-15}J. Burkhardt, R. L.Weaver, Journal of Sound and Vibration 196, 147
(1996); J. Burkhardt, Ultrasonics 36, 471 (1986); O.I. Lobkis, R.L.
Weaver, I. Rozhkov, Journal of Sound and Vibration 237, 281 (2000);
M R Schroeder, 5th Int. Congress of Acoustics p G31 (1965).
\bibitem{key-11}D. V. Savin, V. V. Sokolov, Phys. Rev. E 56 (1997); H.-J. Sommers,
D. V. Savin, V. V. Sokolov, Phys. Rev. Lett. 87, 094101 (2001); D.
V. Savin, H.-J. Sommers, cond-mat/0303083 v2 (2003); P. W. Brouwer,
K. M. Frahm, and C. W. J. Beenakker, Phys. Rev. Lett. 78, 4737 (1997);
Waves Random Media 9, 91 (1999); M.G.A. Crawford, P.W. Brouwer, Phys.
Rev E 65, 026221 (2002).
\bibitem{key-14}H. Alt, H.-D. Gräff et al, Phys. Rev. Lett. 74, 62, (1995).
\bibitem{key-16}K. F. Graff, Wave Motion in Elastic Solids, Dover, NY 1975.
\bibitem{key-17}R. L. Weaver, J. Acoust. Soc. Am. 101, 1441-49 (1997).
\bibitem{key-18}A. D. Pierce, V. Sparrow and K. Russell, J. Vib. Acoust. 117 339-348
(1995).\end{thebibliography}
\end{document}